\documentclass[twocolumn, doublespace, 12pt fonts]{aastex6}
\usepackage{amsmath}

\shorttitle{Constrain NS's EoS}
\shortauthors{Du et al.}

\begin{document}

\title{Constraining the Equation of State of Neutron Stars through GRB X-Ray Plateaus}

\author{Shuang Du$^{1,2}$, Enping Zhou$^{3,1}$, Renxin Xu$^{1}$}
\affil{
$^{1}${State Key Laboratory of Nuclear Physics and Technology, School of Physics, Peking University, Beijing 100871, China}\\
$^{2}${Department of Physics and Astronomy, Sun Yat-Sen University, Zhuhai 519082, China}\\
$^{3}${Max Planck Institute for Gravitational Physics (Albert Einstein Institute), Am M\"{u}hlenberg 1, Potsdam-Golm, 14476, Germany}}

\email{dushuang@pku.edu.cn, r.x.xu@pku.edu.cn}
\begin{abstract}
The unknown equation of state (EoS) of neutron stars (NSs) is puzzling because of rich non-perturbative effects of strong interaction there.
A method to constrain the EoS by using the detected X-ray plateaus of gamma-ray bursts (GRBs) is proposed in this paper.
Observations show some GRB X-ray plateaus may be powered by strongly magnetized millisecond NSs.
The properties of these NSs should then satisfy: (i) the spin-down luminosity of these NSs should be brighter than the observed luminosity of the X-ray plateaus;
(ii) the total rotational energy of these NSs should be larger than the total energy of the X-ray plateaus.
Through the case study of GRB 170714A, the moment of inertia of the associated NS is constrained as $I>1.0\times 10^{45}\left ( \frac{P_{\rm cri}}{1\;\rm ms} \right )^{2} \;\rm g\cdot cm^{2}$,
where $P_{\rm cri}$ is the critical rotational period that an NS can achieve. The constraint of the radii of NSs according to GRB 080607 is shown in Table 1.
\end{abstract}


\section{Introduction}\label{sec1}
Determining the equation of state (EoS) of neutron stars (NSs) is very important for the cognition of low-energy strong interaction \citep{1992PhRvD..46.1274G,2005PrPNP..54..193W,2018SCPMA..61j9531X}.
Common approaches to constrain the EoS are measuring the mass and radius of an NS, or making constraints on the
maximum mass of non-rotating NSs. Our knowledge about the EoS of dense matter has been greatly improved by the recent observation of gravitational wave (GW) radiation from a binary neutron star (BNS) merger (GW170817, \citealt{Abbott2017gw}) and its electromagnetic counterparts (\citealt{2017ApJ...848L..13A}, cf. a recent review \citealt{Baiotti2019}). The fate of the merger remnant is tightly related to the
properties of NS EoSs, and it is with regret that the post-merger remnant of GW170817 is undetermined \citep{2019ApJ...875..160A}.
Although, the total mass of the NS binary $\sim 2.7\;\rm M_{\odot}$ \citep{2017ApJ...848L..13A} is much larger than the measured masses of galactic pulsars,
the EoS with a higher upper limit on rest mass as high as $3.0 \;\rm M_{\odot}$ is still not ruled out (e.g., strangeon star, see \citealt{2019EPJA..55....60L}; \citealt{2018ApJ...861..114Y,2019MNRAS.483.1912P}).

On the other hand, the radii of NSs are so small that it is hard to be measured accurately either through the observation of electromagnetic waves or gravitational waves until now.
By modeling periodic brightness oscillations of `hotspots' on spinning pulsars, some NSs radii can be constrained (see \citealt{2016ARA&A..54..401O} for review).
For example, assuming the NS with a mass of $1.4 \;\rm M_{\odot}$,
the analysis of PSR J0437-4715 leads to the radius $R(1.4\;\rm M_{\odot})\in (6.8-13.8)\;\rm km$ ($90\%$ confidence level; \citealt{2007ApJ...670..668B}).
Similarly, an analysis of NS binary merger GW 170817 whose tidal deformability $\Lambda(1.4\;\rm M_{\odot})<800$ yields
$R(1.4\;\rm M_{\odot})\in (9.9- 13.6)\;\rm km$ ($90\%$ confidence level; \citealt{2018PhRvL.120q2703A}).
Therefore other observations related to BNS merger
remnants are valuable as independent constraints on EoS models.


In this paper, we propose a new approach to constrain the EoS of NSs through the X-ray plateaus of gamma-ray bursts (GRBs).
GRBs can be classified into two categories based on duration $T_{\rm 90}$  \citep{1993ApJ...413L.101K} that short GRBs with $T_{\rm 90}<2\;\rm s$
(originated from double NS mergers, \citealt{2017ApJ...848L..13A})
and long GRBs with  $T_{\rm 90}>2\;\rm s$ (originated from massive star collapses, \citealt{1999ApJ...524..262M}).
Short or long-lived rotationally supported NSs may be born in GRBs \citep{1992Natur.357..472U,1992ApJ...392L...9D,1998PhRvL..81.4301D}
which depends on the remanet masses of these catastrophes.
In principle, the nascent NSs, both conventional NSs and strange stars, can be strongly magnetized (so-called magnetar) through
differential-rotation-induced or convection-induced turbulent dynamo process \citep{1992ApJ...392L...9D,2001A&A...371..963X,2019arXiv190703685T}.
If the central object of a GRB is a millisecond magnetar,
the energy injection \citep{1998A&A...333L..87D,2001ApJ...552L..35Z} by spin-down wind of the magnetar may result in
an X-ray plateau followed by a power-law decay with index $\sim -2$ .
This theoretical expectation is really observed by \emph{Swift }XRT \citep{2009MNRAS.397.1177E} and further strengthen by a recent observation
that the measured light curve of X-ray transient CDF-S XT2 is consistent with the plateau predicted by the millisecond magnetar born in an NS binary merger \citep{2019Natur.568..198X}.
Besides, some X-ray plateaus can be followed by very steep decays (with index $<-3$, the so-called `internal plateau', see e.g., \citealt{2007ApJ...665..599T}).
This feature can be reasonably explained under the magnetar scenario (see \citealt{2015PhR...561....1K} for review).
The spin-down wind of the supramassive magnetar powers the X-ray internal plateau.
The transition from the supramassive magnetar to the black hole through gravitational collapse after losing rotation energy may naturally account for the steep decay phenomenologically.

The luminosities and durations of these X-ray plateaus (case 1: plateau $+$ a decay with index $\sim -2$;
case 2: plateau $+$ a steep decay with index $< -3$) are
closely related to the properties of the central NSs \citep{2016MNRAS.462.2990D,2019MNRAS.482.2973D}.
So through analyzing the relevant observation data,
one may, in turn, constrain the properties of these NSs, such as the EoS.
We describe our method in Section 2. In section 3, two case studies are shown.
In section 4, the angular distribution of the spin-down wind is discussed.
Section 5 is the summary.

\section{The method}
The X-ray plateau is powered by the approximately isotropic spin-down wind of the central magnetar (see e.g., \citealt{2001A&A...369..694S};
we discuss the angular distribution of the spin-down flux in section 4),
such that the spin-down luminosity $L_{\rm sd}$ should be larger than the luminosity of the X-ray plateau $L_{\rm X,pla}$, i.e.,
\begin{eqnarray}\label{Eq1}
L_{\rm sd}=\frac{8\pi^{4}B_{\rm eff}^{2}R^{6}}{3c^{3}P^{4}}>L_{\rm X,pla},
\end{eqnarray}
where $R$ is the equatorial radius, $P$ is the NS period, $B_{\rm eff}$ is the effective dipole magnetic field strength on the NS surface (perpendicular to the rotation axis of the NS),
 and $c$ is the speed of light.
On the other hand, the initial total rotational energy $E_{\rm k,0}$ of the magnetar should be high enough to power the whole X-ray plateau, so one has
( the $=$ is for the case 1, the $>$ is for the case 2)
\begin{eqnarray}\label{Eq2}
E_{\rm k,0}/L_{\rm sd}\geq t_{\rm b},
\end{eqnarray}
where
\begin{eqnarray}\label{Eq4}
E_{\rm k,0}= 2\pi^{2}I/P_{0}^{2},
\end{eqnarray}
$P_{0}$ is the initial period, $t_{\rm b}$ is the break time of the X-ray plateau, and $I$ is the moment of inertia of the magnetar.

According to equations (1)-(3), one has
\begin{eqnarray}\label{Eq5}
P_{\rm cri}<P_{0}< \left ( \frac{2\pi^{2}I }{t_{\rm b}L_{\rm X,pla}} \right )^{1/2},
\end{eqnarray}
\begin{eqnarray}\label{Eq6}
I> \frac{P_{\rm cri}^{2}}{2\pi^{2}} \int_{0}^{t_{\rm b}} L_{\rm X,pla}\; dt,
\end{eqnarray}
where $P_{\rm cri}$ is the critical period that NSs can achieve.

For a certain EoS, given a magnetar mass $M_{\rm mag}$, one can calculate the theoretical values of the radius $R_{\rm th}$,
rotational inertia $I_{\rm th}$, and $P_{\rm cri}$ numerically (see, e.g., \citealt{1992ApJ...390..541W}).
Since $t_{\rm b}$, $L_{\rm X,pla}$ are measurable quantities, the constraint of $I_{\rm th}$ (i.e., equation \ref{Eq5}) is less model-dependent.
Besides, the masses of the central NSs $M_{\rm mag}$ can be roughly classified as three types:
\begin{itemize}
\item[(i)] case 1+short GRB: $2.2\;\rm M_{\odot}<$${M_{\rm mag} }< $$\sim M_{\rm ToV}$, where $M_{\rm ToV}$ is the maximum NS mass for a nonrotating NS.
The lower limit is inferred form the newest observation that the mass of PSR J0740+6620 is $2.17\pm 0.11\;\rm M_{\odot}$ ($68\%$ confidence level, \citealt{2019arXiv190406759C}).
More strictly, the lower limit can be taken as the current record $\sim 2.0\;\rm M_{\odot}$ \citep{2013Sci...340..448A};
\item[(ii)] case 1+long GRB: $M_{\rm mag}\sim 1.4\;\rm M_{\odot}$, since the mass of the individual galactic NS is around $1.4\;\rm M_{\odot}$ \citep{2016ARA&A..54..401O};
\item[(iii)] case 2: $M_{\rm max}<M_{\rm mag}<1.3M_{\rm max}$ , since if $M_{\rm mag}$ is larger than $1.3$ times of the maximum mass of a rotating NS $M_{\rm max}$,
 the nascent NS will collapse to a black hole during its dynamical timescale \citep{2008PhRvD..78h4033B,2011PhRvD..83l4008H}. We already know that  $M_{\rm max}$ is either greater than $2.7\;\rm M_{\odot}$ or less than $2.7\;\rm M_{\odot}$ \citep{2017ApJ...848L..13A}.
\end{itemize}
For the magnetars under these three types, $I$ should be also consistent with the mass range.

The constraint of $R$  can not be as rigorous as $I$.
If magnetars do exist in the case 1 and case 2, there is at least one pair of parameters $(B_{\rm eff, max}, P_{\rm cri})$ that makes equation (1) work, i.e.,
\begin{eqnarray}\label{Eq7}
R&&>6.9\times 10^{5}\left ( \frac{P_{\rm cri}}{1\;\rm ms} \right )^{1/3}\left ( \frac{B_{\rm eff,max}}{10^{15}\;\rm Gs} \right )^{-1/3}\nonumber\\
&&\;\;\;\times\left ( \frac{L_{\rm X,pla}}{10^{48}\;\rm erg\cdot s^{-1}} \right )^{1/6}\; \rm cm.
\end{eqnarray}
In principle, according to the dynamo mechanism,
$B_{\rm eff}$ has a upper limit \citep{1992ApJ...392L...9D} that
\begin{eqnarray}\label{Eq8}
B_{\rm eff,max}=3\times 10^{17} (P/1\;\rm ms)^{-1}\;\rm G.
\end{eqnarray}
But the observations of soft gamma repeaters (SGRs) and anomalous X-ray pulsars (AXPs) show that almost
all the associated magnetars have periods $P_{\rm t}\in (1- 10)\;\rm s$, time derivative of period $\dot{P}_{\rm t}\in (10^{-11}- 10^{-10})\;\rm s\cdot s^{-1}$
and inferred magnetic fields $B_{\rm eff}\in (10^{14}- 10^{15})\;\rm Gs$ (except the uncertain magnetic field strength of SGR 1806-20,
whose upper limit is perhaps as high as $2.5\times 10^{15}\;\rm Gs$; see \citealt{2007ApJ...654..470W}).
The existences of the X-ray plateaus show that the period of the nascent magnetars are $\sim 1\rm ms$ \citep{2013MNRAS.430.1061R,2016MNRAS.462.2990D}.
If this is also true for the magnetars in SGRs and AXPs, through the assumption that the magnetic moment
of these magnetars does not change significantly, one has the ages of these magnetars, i.e.,
\begin{eqnarray}\label{Eq9}
\tau\approx\frac{P_{\rm t}}{2\dot{P}_{\rm t}}\sim 10^{4}\;\rm yr.
\end{eqnarray}
So these magnetars are young NSs which is consistent with the observation and model of SGRs (e.g., \citealt{1982ApJ...255L..45C,2016ApJ...826..226K}).
Besides, if the decay of magnetic torque of a magnetar is consistent with the galactic pulsars that the decay time scale of the magnetic torque $\tau_{\rm D}\sim (10^{6}- 10^{7})\;\rm yr$ \citep{1975MNRAS.171..579L},
one can see $\tau\ll \tau_{\rm D}$, such that the assumption of quasi-constant  magnetic torque is reasonable.
Hereinafter,  we take the upper limit of $B_{\rm eff}$ as $B_{\rm eff, max}=10^{15}\;\rm Gs$ and $B_{\rm eff, max}=2.5\times 10^{15}\;\rm Gs$ empirically.

In the above discussion, the gravitational radiation of magnetars is ignored for conservatism. Because if taking the gravitational radiation into consideration,
the constraint equations (\ref{Eq6}) and (\ref{Eq7}) will be more tighter since the required $E_{\rm k,0}$ is larger.
Actually, whether the spin down of a sample is dominated by gravitational radiation can be inferred from the X-ray afterglow.
From conservation of energy, the spin down of NS is generally read as  \citep{1983bhwd.book.....S}
\begin{eqnarray}\label{Eq13}
I\Omega \dot{\Omega }=-L_{\rm em}-L_{\rm gw}=-\frac{B_{\rm eff}^{2}R^{6}\Omega ^{4}}{6c^{3}}-\frac{32GI^{2}\epsilon ^{2}\Omega ^{6}}{5c^{5}},
\end{eqnarray}
where $\epsilon$  and $\Omega=2\pi/P$ are the ellipticity and angular velocity of the NS respectively, and the over dot is time derivative.
The asymptotic solution of equation (\ref{Eq13}) can be solved as follows \citep{2016MNRAS.458.1660L}.
When the magnetic dipole radiation dominates the spin down, there is
\begin{eqnarray}\label{Eq14}
L_{\rm sd}(t)=L_{\rm sd,0}\left ( 1+\frac{t}{\tau_{\rm em}} \right )^{-2},
\end{eqnarray}
where
\begin{eqnarray}\label{Eq17}
\tau_{\rm em}=\frac{3c^{3}I}{B_{\rm eff}^{2}R^{6}\Omega_{0}^{2} } ,
\end{eqnarray}
$L_{\rm sd,0}$ and $\Omega_{0}$ are the luminosity and angular velocity at $t = 0$.
When the spin down is dominated by gravitational radiation, there is
\begin{eqnarray}
L_{\rm sd}=L_{\rm sd,0}\left ( 1+\frac{t}{\tau _{\rm gw}} \right )^{-1},
\end{eqnarray}
where
\begin{eqnarray}\label{Eq16}
\tau_{\rm gw}=\left ( \frac{5c^{5}}{128GI\epsilon ^{2}\Omega _{0}^{4}} \right ).
\end{eqnarray}
Note that the decay of gravitational radiation $(L_{\rm gw}\propto \Omega^{6})$ is faster than the decay of  magnetic dipole radiation $(L_{\rm em}\propto \Omega^{4})$.
Once the spin down is dominated by the gravitational radiation initially, there will be a moment $\tau_{\ast}$ that the spin-down luminosity changes from being
gravitational radiation dominated to being dominated by magnetic dipole radiation \citep{2001ApJ...552L..35Z,2016MNRAS.458.1660L}, i.e.,
\begin{eqnarray}\label{Eq15}
\tau_{\ast}=\frac{\tau_{\rm em}}{\tau_{\rm gw}}\left ( \tau_{\rm em}-2\tau_{\rm gw} \right ).
\end{eqnarray}
So the decay index of the X-ray afterglow will be changed as $\sim 0\rightarrow \sim -1\rightarrow \sim -2$ (a similar result based on NS $r$-mode instability and several candidates
can be seen in \citealt{2010ApJ...715..477Y};
the case studies in section 3 are not in this situation).
Besides, there is another interesting issue.
For a given EoS, one roughly has $R$ and $I$. Through the effect of the gravitational radiation on the slope of X-ray light curves,
$\tau_{\rm gw}$ and $\tau_{\ast}$ is measurable in some cases.
Combining equations (\ref{Eq1}), (\ref{Eq17}), (\ref{Eq16}) and (\ref{Eq15}), one can estimate $\epsilon$.


\section{Case studies}
Until now we have not found a extreme sample that can make a tight limit on $I$.
For example, for GRB 170714A whose total energy of the X-ray plateau is (in $0.3-10\;\rm keV$, the luminosity of the X-ray plateau is from \citealt{2018ApJ...854..104H};
different with \citealt{2018ApJ...854..104H}, we assume the two plateaus of GRB 170714A are all powered by spin-down wind.)
\begin{eqnarray}
E_{\rm X,pla}=\int_{0}^{t_{\rm b}}L_{\rm X,pla}\; dt\approx 2.0\times 10^{52}\;\rm erg.
\end{eqnarray}
Through equation (5), one has
\begin{eqnarray}\label{Eq10}
I>1.0\times 10^{45}\left ( \frac{P_{\rm cri}}{1\;\rm ms} \right )^{2} \;\rm g\cdot cm^{2}.
\end{eqnarray}
Almost all the EoSs can match this result. But it is worth emphasizing that the spin-down energy
of magnetar will not be completely converted to X-ray emission, and
$E_{\rm X,pla}=2.0\times 10^{52}  \;\rm erg$ is just the energy in $(0.3-10)\;\rm keV$. Considering these two factors,
the value of $E_{\rm X,pla}/E_{\rm k,0}$ should be appropriately less than $1$.
In the future, if detectors can give a wider energy-band observation
to GRB X-ray plateaus, the constraint of equation (\ref{Eq10}) will be tighter.

 \begin{figure}
    \centering
    \includegraphics[width=0.45\textwidth]{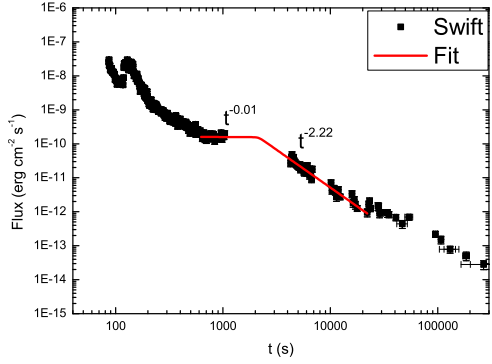}
    \caption{The fitting result of the X-ray afterglow of GRB 080607. One can see that the magnetar candidate belongs to type (\rm ii).}
    \label{wolf}
    \end{figure}

A good sample to constrain $R$ is that a brighter X-ray plateau which meets case 1 or case 2.
As an example, a magnetar candidate that GRB 080607 is used to be a case study.
GRB 080607 is a long GRB with $T_{90}=79\;\rm s$ \footnote{Stamatikos et al., GCN 7852, https://gcn.gsfc.nasa.gov/gcn3/7852.gcn3},
and redshift $z_{0}=3.04$ \footnote{Prochaska et al., GCN 7849, https://gcn.gsfc.nasa.gov/gcn3/7849.gcn3}.
We fit the X-ray afterglow of  GRB 080607  with a smooth broken power law that the break time of the plateau is $t_{\rm b}\approx 2200\;\rm s$,
the decay index before the break is $\alpha_{1}\approx -0.01$ and the decay index after the break is $\alpha_{2}\approx -2.22$  (see Fig. 1).
One can find that the magnetar candidate of GRB 080607 belongs to the type ($\rm ii$), such that the mass of the magnetar may be around $1.4\rm \; M_{\odot}$.
The mean unabsorbed flux of the X-ray plateau in $(0.3-10)\;\rm keV$ is $F_{\rm lux}=2.26_{ -0.19}^{+0.22}\times 10^{-10}\;\rm erg\; cm^{-2}\; s^{-1}$ \citep{2009MNRAS.397.1177E}.
Adopting $\rm \Lambda CDM$ model with $H_{0}=70\;\rm km\; s^{-1}\; Mpc^{-1}$, $\Omega_{\rm m}=0.3$ and $\Omega_{\rm \Lambda}=0.7$, one has
\begin{eqnarray}\label{Eq11}
L_{\rm X,pla}(080607)=4\pi D_{\rm L}^{2} F_{\rm lux}>1.6\times 10^{49}\;\rm erg \;s^{-1},
\end{eqnarray}
where $D_{\rm L}$ is the luminosity distance from the source to the earth.
Through equation (\ref{Eq7}), the constraint of $R$ is shown in top half part of Table 1 \footnote{Note that when the rotation of a NS is near to disintegration, i.e.,
$P\sim P_{\rm cri}$, $B_{\rm eff}$ should be small enough to keep $L_{\rm sd}$ below the jet power.}.
In view of that different EoSs have different critical periods, we adopt two different values of $P_{\rm cri}$.
The constraint by equation (\ref{Eq7}) is conservative, since the shorter the period, the weaker the constraint of $R$.
Note the newborn magnetar may have a fast decay but intense GW radiation due to some instabilities
 (e.g., \citealt{1998ApJ...502..708A,1998PhRvD..58h4020O}; see \citealt{2009CQGra..26f3001O} for review),
even differential rotation \citep{2019arXiv190209361Z},
we also consider the situation that the initial period of the magnetar is several times the critical period (see the latter part of Table 1) .

\begin{table}[!htbp]
\caption{The constraint of equatorial radius}
\begin{center}
\begin{tabular*}{6cm}{lll}
\hline
$B_{\rm eff,max}\;(10^{15}\rm Gs)$ & $P_{\rm cri}\;(\rm ms)$  &$R\; (10^{5}\rm cm)$ \\
\hline
\;\;$1.0$  & 0.5 &$>8.7$\\
\;\;$1.0$  & 1.0 &$>11.0$ \\
\;\;$2.5$  & 0.5 &$>6.4$ \\
\;\;$2.5$  & 1.0 &$>8.1$\\
\hline
$B_{\rm eff,max}\;(10^{15}\rm Gs)$ & $P_{0}\;(\rm ms)$  &$R\; (10^{5}\rm cm)$ \\
\hline
\;\;$1.0$  & 1.5 &$>12.5$\\
\;\;$2.5$  & 1.5 &$>9.2$\\
\;\;$1.0$  & 2.0 &$>13.8$\\
\;\;$2.5$  & 2.0 &$>10.2$\\
\hline
\end{tabular*}
\end{center}
\end{table}

\section{angular distribution}
 \begin{figure}[h]
    \centering
    \includegraphics[width=0.4\textwidth]{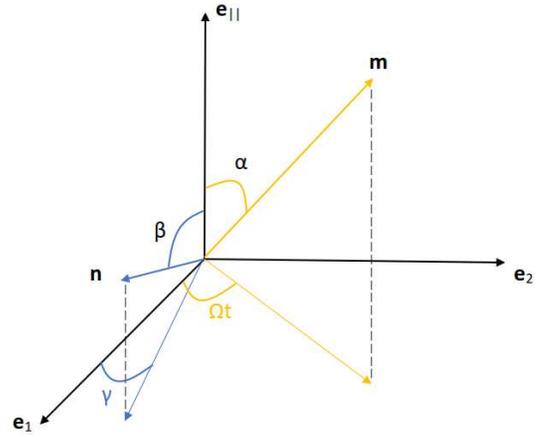}
    \caption{Schematic diagram of the geometric relationship of vectors.
    $\mathbf{e}_{\parallel}$ is a unit vector parallel to the rotation axis, $\mathbf{e}_{1}$ and $\mathbf{e}_{2}$ are the orthogonal unit vectors perpendicular to rotational axis.}
    \label{geo}
    \end{figure}
According to electrodynamics \citep{1975ctf..book.....L}, the energy flux of the pure magnetic dipole field is
\begin{eqnarray}
\mathbf{S}&=&\frac{1}{4\pi c^{3}d^{2}}|(\ddot{\mathbf{m}}\times \mathbf{n})\times\mathbf{n}|^{2}\mathbf{n}\nonumber\\
&=&\frac{1}{4\pi c^{3}d^{2}}|\ddot{\mathbf{m}}|^{2}\mathbf{n}\sin^{2}\theta,
\end{eqnarray}
where $\mathbf{\mathbf{n}}$ is the unit vector from the source to observer, $\mathbf{m}$ is magnetic moment, $d$ is the distance from the source to observer,
and $\theta$ is the angle between $\mathbf{n}$ and $\mathbf{m}$.
Through the geometry shown in Fig. 2, there is
\begin{eqnarray}
\cos\theta=&&(\cos\beta\cos\alpha+\sin\beta\cos\gamma\sin\alpha\cos\Omega t\nonumber\\
&&+\sin\beta\sin\gamma\sin\alpha\sin\Omega t).
\end{eqnarray}
Then the period average of $\mathbf{S}$ is
\begin{eqnarray}\label{Eq18}
\bar{S}=\frac{1}{4\pi c^{3}d^{2}}|\ddot{\mathbf{m}}|^{2}(1-\cos^{2}\alpha\cos^{2}\beta-\frac{1}{2}\sin^{2}\alpha\sin^{2}\beta),
\end{eqnarray}
According to equation (\ref{Eq18}), there is a critical state that $\mathbf{n}\parallel \mathbf{e}_{\parallel}$, and $\mathbf{n}\perp \mathbf{m}$.
Therefore, for the NS far away form the earth, the flux observed on the earth satisfies
\begin{eqnarray}
F_{\rm lux}\leq \bar{S}_{\rm max}= \frac{1}{4\pi c^{3}D_{\rm L}^{2}} |\ddot{\mathbf{m}}|^{2}.
\end{eqnarray}
Correspondingly, the luminosity of magnetic dipole radiation satisfies
\begin{eqnarray}
L_{\rm em}&&= \frac{2}{3c^{3}}|\ddot{\mathbf{m}}|^{2}=\frac{8\pi}{3}D_{\rm L}^{2}\bar{S}_{\rm max},
\end{eqnarray}
and then there is
\begin{eqnarray}
L_{\rm X,pla}\geq \frac{2}{3}\times 4\pi D_{\rm L}^{2}F_{\rm lux}.
\end{eqnarray}
To be more more stringent, the $L_{\rm X,pla}$ in equation (\ref{Eq1}) should be divided by $3/2$.
However this angular correction may be offset by the fact that $E_{\rm X,pla}/E_{\rm k,0}<1$, we do not consider it in the above discussion.

\section{Summary }
In this paper, we aim to propose a new approach to constrain the EoS of NSs.
The method to constrain the rotational inertia $I$ is less model-dependent, but the constraint of the radius $R$ is somewhat empirical.
Till now, we do not find any perfect samples which can constrain the EoS tightly (similar to the method described in introduction).
Two case studies of GRB 170714A and GRB 080607, and the constraints of $I$ and $R$, are shown in section 3.

Nevertheless, it should be kept in mind that the radius constraint we put with this method (i.e., the values shown in Table 1) is on the equatorial radius of a massive rotating neutron star formed either by the collapse of a massive star (in the case of long GRB) or a BNS merger (in the case of short GRB). Therefore, it should not be directly compared with other constraints, such as GW170817 constraint (i.e., $R(1.4\;\rm M_{\odot})\in(9.9-13.6)\;\rm km$) which is on the radius of a non-rotating $1.4\;\rm M_{\odot}$ NS. However, in the future, those constraints can be combined together if an X-ray plateau observation is achieved together with the GW observation of a BNS merger. In addition to $R(1.4\;\rm M_{\odot})$ or $\Lambda(1.4\;\rm M_{\odot})$,
the GW observation will provide information about the mass and even spin frequency (if post-merger GW signal could be obtained) of the remnant magnetar. Meanwhile, our method can provide constraints to the rotational inertia and equatorial radius. On the premise that our method is correct, any EoS model should satisfy these constraints through GW observationas as well as the constraints imposed by the associated X-ray plateaus at the same time.


To improve the method described in section 2, one can consider the angular distribution of the spin-down winds of NSs (e.g. section 4) and
the relativistic modification on the rotational energy of NSs (through numerical relativity).
In order to give a better constraint, there are two ways to improve:
(a) widening the observational energy band, e.g., $(0.1- 30)\;\rm keV$ or even more higher which depends on the hard X-rays telescopes (e.g., HXMT, \citealt{2014SPIE.9144E..21Z}; eXTP, \citealt{2016SPIE.9905E..1QZ});
(b) searching for some extreme samples with long-duration and bright X-ray plateaus.


\section{Acknowledgement}
We thank the anonymous referee for helpful comments.
We acknowledge the use of the public data from the \emph{Swift} data archives.
We thank Fang-Kun Peng for the useful discussion about the estimation of ellipticities of NSs during
973 programme annual meeting ``gamma-ray bursts and related frontier physics''  in Kunming on January 2018.
This work was supported by the National Key R\&D Program of China (Grant
No. $\rm 2017YFA0402602$), the National Natural Science Foundation of China
(Grant Nos. $11673002$, and $\rm U1531243$), and the Strategic Priority Research
Program of Chinese Academy Sciences (Grant No. $\rm XDB23010200$).

\end{document}